\documentclass[12pt,preprint]{aastex}
\shorttitle{The Hot Component of AG Dra} 
\shortauthors{Sion et al.} 

\begin{document}

\title{On the Nature of the Hot Component in the Symbiotic, 
Supersoft X-ray Binary AG Draconis\altaffilmark{1}}

\author{Edward M. Sion, Jackeline Moreno, Patrick Godon\altaffilmark{2}} 
\affil{Astronomy \& Astrophysics, Villanova University, \\
800 Lancaster Avenue, Villanova, PA 19085, USA}
\email{
edward.sion@villanova.edu, 
jackeline.moreno@villanova.edu,
patrick.godon@villanova.edu} 
\author{Bassem Sabra} 
\affil{Dept. of Physics \& Astronomy, Notre Dame University  Louaize; Zouk Mosbeh, Lebanon}\email{bsabra@ndu.edu.lb}\and 
\author{Joanna Mikolajewska}
\affil{Copernicus Astronomical Center\\  Warsaw, Poland}
\email{mikolaj@camk.edu.pl}  

\altaffiltext{1}{Based on observations made with the NASA-CNES-CSA 
{\it Far Ultraviolet Spectroscopic Explorer (FUSE)} was operated 
for NASA by the Johns Hopkins University under NASA contract 
NAS5-32985. }  

\altaffiltext{2}{Visiting at the Henry Rowland Department of Physics,
The Johns Hopkins University, Baltimore, MD.} 

\begin{abstract} 

AG Dra is a symbiotic variable consisting of a metal poor, yellow giant mass donor under-filling its Roche lobe, and a hot accreting white dwarf, possibly surrounded by an optically thick, bright accretion disk which could be present from wind accretion. We constructed NLTE synthetic spectral models for white dwarf spectra and optically thick accretion disk spectra to model a FUSE spectrum of AG Dra, obtained when the hot component is viewed in front of the yellow giant. The spectrum has been de-reddened (E(B-V) = 0.05) and the model fitting carried out, with the distance regarded as a free parameter, but required to be larger than the Hipparcos lower limit of 1 kpc. We find that the best-fitting model is a bare accreting white dwarf with $M_{wd}= 0.4 M_{\odot}$, T$_{eff}$ = 80,000K and a model-derived distance of  1543 pc. Higher temperatures are ruled out due to excess flux at the shortest wavelengths while a lower temperature decreases the distance below 1 kpc. Any accretion disk which might be present is a only a minor contributor to the FUV flux. This raises the possibility that the soft X-rays originate from a very hot boundary layer between a putative accretion disk and the accreting star.

\end{abstract}


\section{Introduction}

AG Dra is a D-Type Symbiotic with an orbital period P$_{orb}\sim 550$ days.The mass donor is a metal-deficient K0-4 giant which has also been classified as a Barium star. Several estimates of the mass loss from the donor giant have appeared 
ranging from $10^{-7}$ M$_{\sun}/yr$ to $2.5 \times 10^{-7}$ M$_{\sun}/yr$ 
(cf. Mikolajewska et al. 1995; Tomov et al. 2000; Tomov \& Tomov 2002). A key 
question is how much of the outflow from the cool giant is accreted 
onto the hot component. The hot component is presumed to be a white 
dwarf though no direct detection has been reported, and it remains 
unclear whether there is an accretion disk (from wind accretion) 
surrounding the hot component. The outburst recurrence time of 
AG Dra is every $\sim 15$ years and the duration of the outbursts is 
$\sim 3 - 6$ years. Hipparcos has yielded a lower limit to the distance 
of $d > 1$ kpc. Mikolajewska et al's (1995) distance  of 2.5 kpc 
can be compared with distance estimates based upon estimates for 
the cool giant radius from its rotational velocity. In particular, 
Fekel et al. (2003) derived $v_{rot} \sin{i}= 3.6 \pm 1.0$ km/s. 
Assuming corotation with the orbital period and using the inclination 
derived from the spectropolarimetry,  an estimate of the red giant radius, 
$R_g = 0.209$(+0.151/-0.075) a.u., is obtained, which combined with the 
near-IR K-magnitude = 6.2, corresponds to $d=1.5 (\pm 0.5)$kpc. 
The errors are set by the combined errors of the rotational velocity 
and inclination estimates. It also appears (Mikolajewska et al. 1995; 
Smith et al. 1996) that the giant donor star in AG Dra is brighter 
than typical for K III giant stars of solar metallicity.

AG Dra is not an eclipsing symbiotic system, therefore the orbital inclination is below $i = 80 \degr$ and estimates between $i = 70 \degr$ and $i = 40 \degr$ have appeared in the literature by Mikolajewska et al. (1995) and Schmid \& Schild 1997), with the latter authors' estimate based on spectropolarimetry of the Raman scattered O VI lines. In our model fitting described in section 3, we ruled out $i = 60 \degr$ and adopted $i = 41 \degr$. The reddening has been estimated to be E(B-V) = 0.05 (Mikolawjewska et al. 1995), however, reddening  
values as high as E(B-V) = 0.10 have not been ruled out (Young et al. 1995). AG Dra has  been classified as a supersoft X-ray source (Greiner et al. 1997). It has been  established that the X-ray/FUV flux level of AG Dra is anti-correlated with the optical maxima (Greiner et al. 1997).

The mass of the hot component has been estimated to be low.  Mikolajewska et al. (1995) found M$_{h}$ = 0.4 to 0.6 M$_{\sun}$ for a distance of 2.5kpc, while Tomov et al. (2000) gives $0.5$ M$_{\sun}$. The hot component is presumably a white dwarf though Mikolajewska et al. (1995) suggests it is a sub-dwarf with $R_{wd} \approx 0.06-0.08R_{\odot}$. Note that a low mass for AG Dra's hot component would be consistent with the finding of Muerset et al.(1991) that the masses of the hot components in symbiotic systems tend to be low. 

However, all previous determinations of the temperature of the hot component have been carried out primarily with either blackbody fits to the X-ray and IUE ultraviolet data or by the application of the modified Zanstra method, using the emission line strengths. An Analysis of the X-ray data has yielded $\sim 160,000$K $\pm$20,000K 
from ROSAT data (Greiner et al. 1997) while Mikolajewska et al.(1995) found T$_{h}$ =  80,000K to 150,000K from blackbody fits to the IUE FUV data. Gonzalez-Riestra et al.(1999) found an average Zanstra temperature of 109,600K $\pm $5400K. 
Likewise, estimates of the accretion rate have been made using IUE data by 
Mikolajewska et al.(1995), Gonzalez-Riestra et al. (1999) during quiescence, Viotti et al. (1999) used the N V (1240) emission measure 
and by Tomov et al.(2000).

The acquisition of FUSE spectra by Young et al.(1995) and X-ray data with XMM-Newton (Gonzalez-Riestra et al.2008) has offered deepened insight into the hot component. There has never been, to our knowledge, a direct comparison between
the FUSE spectrum of AG Dra and state of the art models of optically thick accretion disks with vertical structure or with NLTE white dwarf photospheric models. Moreover, even if the continuum (and absorption line) radiation in AG Dra's FUV spectra does not arise from an accretion disk, it is just as important to determine the hot star properties by direct comparison of the FUV spectra with NLTE high gravity photospheres. This modeling would offer an important comparison with indirect methods (e.g. modified Zanstra method) or black body fits to the FUV spectra. Why is this comparison important? Because it remains unclear at this time what comprises the hot component in AG Dra. Is it the inner edge of what may be an accretion disk or the surface of a bloated accreting white dwarf? Does the accretion of wind from a giant star underfilling its Roche lobe lead to an accretion disk plus boundary layer formation or not? If a disk is present, then can the spectral energy distribution be fitted by a realistic accretion disk model? Furthermore, the rate of accretion onto the hot component, whether via a disk or by direct gravitational capture of wind outflow from the donor giant, holds the key to understanding the mechanism of AG Dra's two kinds of outbursts, as well as its intervals of low activity and quiescence.

In section 2, we present the archival FUSE data for AG Dra.

\section{Spectroscopic Observations}   

The FUSE spectrum (see Young et al. 1995) of AG Dra near phase 0.5 (S312010200) was obtained on 03-16-2000 with a start time of 15:57:00 UT. The spectrum was taken with LWRS with an exposure time of 2388 seconds.  The identified lines are shown in Fig. 1. The FUSE emission lines were discussed at length in Young et al. (1995).   
When the FUSE spectrum was obtained, the white dwarf was being viewed in front of the giant companion and during a low brightness state of AG Dra. We de-reddened the spectrum with E(B-V) = 0.05.

In preparation for the fitting of the FUSE spectrum at 0.5 phase, 
we masked all emission lines, as well as ISM absorption features
(see Fig. 3 for details). 

The volume emission measure of the nebular continuum, 
$n_e^2V/4 \pi d^2 \sim 5-10 \times 10^{13}$cm-5 from the Balmer H$\beta$ line, for a reasonable 
electron temperature $Te \sim 10000-20000$K. This emission measure is also consistent with the 
observed mm/submm emission (e.g. Mikolajewska et al. 2003 and 
and references therein).  
We have not included a contribution from the nebular continuum.
By model fitting the spectral range of the FUSE spectrum, any contribution
from a nebular continuum is negligible (Skopal 2005) since both during the quiet phase
and active phase, the nebular continuum contributes only longward of 1200A.

However, we have explored whether the temperature of the hot component is constrained by the nebular emission lines, based upon the assumption implicit in the Zanstra method that they form from photoionization by the hot component. We ran simulations with the code CLOUDY (Ferland 1998) in order to constrain the nature of the source powering the nebulosity in AG Dra (B.Sabra 2012, in preparation). We generated simulated emission line ratios and compared them to the observed line ratios. We set up a grid of simulated nebulae, all with solar abundances, and plotted the fluxes of the resulting emission lines in line ratio diagrams. Every nebula in the grid is illuminated by an incident continuum from a blackbody at a given temperature, and has a given hydrogen density and ionization parameter, defined as the ratio of hydrogen ionizing photons to hydrogen density. For AG Dra, we chose radiation temperatures of 75,000K, 100,000K, and 125,000K to represent the hot component source in AG Dra.  The hydrogen densities values are $\log n_{\rm H}/cm^{-3}=3, 6, 9, 12$. The ionization parameter values are $\log U=-4, -3, -2, -1, 0, 1, 2, 3, 4$. In Fig.2a,b, we display the results of CLOUDY photoionization simulations of AG Dra overplotted with FUSE and IUE observed line ratios as listed in Table 2 of Young et al. (2005), and in Table 4 of Mikolajewska et al. 1995 (JD 4820, phase 1.686 values), respectively. We chose these datasets since they contain a large number of emission lines. Fig.2a (left panel) displays the behavior of the low ionization line ratios while Fig.2b (right panel) displays the high ionization line ratios. The observed line ratios are denoted by "X", while squares, rhombii, and crosses, and result from incident radiation due to a blackbody at 75,000K, 100,000K, and 125,000K, respectively. The change in density has a minimal effect on the simulated line ratios. Its effect is even less than that of the blackbody temperature. The sequence in the line ratios is primarily due to the changing ionization parameter. The change in the blackbody temperature does not result in a substantial change in the simulated line ratios, especially those involving lower ionization lines. The lower ionization lines show better agreement with the observations
than the higher ionization lines. The latter could posssibly also be explained by shock heating. We find that the emission line strengths and their ratios remain fairly constant for given temperatures between 50,000K and 125,000K. Hence, they do not provide a sufficiently accurate and strong constraint on the hot star temperature. 

\section{White Dwarf Atmosphere and Accretion Disk Model Fitting Results}  

We used the NLTE option in TLUSTY (Hubeny 1988) and SYNSPEC (Hubeny and Lanz 1995) to construct 
the NLTE model atmospheres and the grid of optically thick, steady state accretion disk models of Wade and Hubeny (1998) for the disk modeling. The inclination of the system, $i = 41 \degr$,  was kept fixed in the fitting. The mass of the white dwarf is assumed to be in the range $0.4-0.6 M_{\sun}$ (log $g = 7- 8$), which is consistent with the lower masses inferred for the hot components in other symbiotic systems (Muerset et al.1991). We used the mass-radius relation of Wood (1995) from H-rich white dwarf evolutionary models.  Our use of models with white dwarf surface gravities seems more plausible than taking gravities lower than log g = 7, since we expect the accreting star to be fully degenerate with an extended envelope rather than having the radius and gravity of a subdwarf. In the symbiotic literature, the term subdwarf is often used but is a misnomer. The physically correct characterization of the hot component is a high luminosity degenerate star, that is, high luminosity white dwarfs with extended atmospheres and H-burning shells (Sion et al. 1978; Sion \& Starrfield 1994). Thus the use of the term subdwarf implies a lower gravity, semi-degenerate object and should be avoided.
 
The distance, d, from the scale factor (normalization) of a white dwarf model photosphere fit is given by

$d = 1000$ (R$_{wd}$/R$_{\sun}$) / $\sqrt{S}$,  

where S is the scale factor of the fit. The scale factor, normalized to a kiloparsec, can be related to the white dwarf radius through: 
$F_{\lambda(obs)} = 4 \pi (R^{2}/d^{2})H_{\lambda(model)}$. 
The factor $4 \pi$ comes from the conversion of Eddington flux, $H_{\lambda}$, resulting from the program SYNSPEC, to physical flux.

For an accretion disk fit, the models are normalized to 100 pc from Earth.
Therefore, the distance from the scale factor of a model accretion disk fit is

$d = 100$ / $\sqrt{S}$.  

We require any fitting solution to correspond to a model-derived distance greater than 1 kpc, the Hipparcos lower limit to AG Dra's distance. Thus, we regard any model-derived distance $d < 1$ kpc to be implausible. 
Accordingly, we searched for a best fitting solution with a low mass white dwarf and hence a larger radius. The increase in the stellar radius leads to distances greater than 1 kpc, the Hipparcos lower limit. In order to obtain a distance of at least 1000 pc, then the surface temperature cannot be lower than 70,000K. All of the models have solar composition. Of course, the giant component has low metallicity rather than  solar composition but it is unlikely that this composition difference will substantially affect our fitting results. 

Subject to the above parameter constraints, we first carried out model accretion disk fits to the FUSE spectrum. 
We found that with $M_{wd}=0.4 M_{\odot}$, $\dot{M} = 5 \times 10^{-8}$ M$_{\sun}$/yr, and $i = 41 \degr$, reasonable parameters for AG Dra, there is a flux deficit in the model at the shortest FUSE wavelengths, relative to the observed spectrum. When we tried an inclination $i = 60 \degr$, the scale factor-derived distance decreases below 1000 pc. 
If we fix the distance for the FUSE spectrum of AG Dra with the white dwarf in front of the red giant, with the distance of AG Dra fixed at 2.5 kpc, a widely used distance estimate, then an accretion rate $\dot{M} = 5 \times 10^{-8}$ M$_{\sun}$/yr and $i = 41 \degr$, yields a poor fit. 

In view of these results with model accretion disks, we tried fits of single temperature NLTE white dwarf photosphere models to the FUSE spectrum. All of the models have solar composition. For a $0.4 M_{\sun}$ white dwarf fit with T$_{eff}$ = 70,000K, the resulting distance d = 992 pc, just below the Hipparcos lower limit. For a $0.4 M_{\sun}$ white dwarf but with T$_{eff}$ = 100,000K, the model-derived distance is 2 kpc, but there is a model flux excess at the shorter wavelengths of the FUSE spectral range relative to the observed spectrum. We found that the best-fitting model was a $0.4 M_{\sun}$ white dwarf with T$_{eff}$ = 80,000K $\pm $5000K and a model-derived distance of 1543 pc. However, our model photosphere fits to the FUSE spectrum for $0.5 M_{\sun}$ yielded reasonable fits as well but with a closer distance barely consistent with the   Hipparcos lower limit. In Fig.3, we display the three different model fits to the FUSE spectrum for a white dwarf mass of $0.4 M_{\sun}$. The dashed line is the accretion disk fit, the dotted line is the 
white dwarf photosphere fit with T$_{eff}$ = 100,000K and the
solid line is the white dwarf photosphere fit with T$_{eff}$ = 80,000K.  

At these high temperatures ($ > $70,000K), a white dwarf photosphere alone provides virtually all of the FUV radiation
from AG Dra. Hydrogen shell burning driven by accretion may be providing the luminosity of the hot component and may explain the supersoft X-rays. If an accretion disk is present, then it provides only a relatively small fraction of the FUV flux while the hot shell burning accreting white dwarf provides $> 90$\% of the FUV flux. Since the FUSE spectrum was obtained during a low optical brightness state, it is not unexpected that a bare white dwarf provides the best model. 

\section{Discussion and Conclusions} 

The most surprising result of our analysis is that our best estimate 
of the surface temperature of the white dwarf is only 80,000K. 
These high gravity NLTE photosphere and disk model fits are, to our knowledge, the first to be carried out for AG Dra. Our results, from the analysis of a FUSE spectrum,
strongly suggest that the true photospheric temperature of the white 
dwarf in AG Dra is considerably cooler than the 
temperatures derived by modified Zanstra methods. No increased amount of interstellar or circumbinary reddening
could change this conclusion and make the white dwarf as hot as 
indicated by the Zanstra method. It is possible that other mechanisms
than photoionization of the cool star wind (e.g. collision shocks, disk coronae) are contributing to the emission line strengths. The Zanstra method also ignores a hot component wind which could render the equivalent widths of He II and H lines to be an inappropriate measure of the true hot component temperature.  

Because of the (relatively) low mass of the accreting star 
and low mass accretion rate (it was observed during a low state), 
it is {\it very unlikely} that the WD temperature reaches 100,000K. 
However, a hot boundary layer could significantly  
contribute to the FUV continuum if there is a disk. 
For a mass accretion rate of the order of $\sim 10^{-9} 
M_{\odot}$/yr, the boundary layer theory predicts an optically thick boundary layer with a temperature in the range $\sim$100,000K (Godon et al. 1995) 
to $\sim 200,000$K (Popham \& Narayan 1995). 
Such a boundary layer would produce an FUV flux similar to the one
we modeled with a 110,000K WD , therefore possibly bringing the distance
in the fitting closer to the estimated 2.5kpc. At the same time, the boundary 
layer would also explain the observed soft X-rays.   

We thank an anonymous referee for useful comments which have helped us to strengthen the manuscript. 
This work was supported in part by NSF grant AST-0807892 to Villanova University.Support for JM was provided by the Polish National Science Center grant no. DEC-2011/01/B/ST9/06145.  
Support was also provided in part by a summer undergraduate research fellowship
from the NASA- Delaware Space Grant Consortium.
Additional support was provided by the National Aeronautics and Space Administration
(NASA) under grant number NNX08AJ39G issued through the Office of Astrophysics Data
Analysis Program (ADP) to Villanova University. PG wishes to thank William P. Blair at the Henry Augustus Rowland Deparment of
Physics \&  Astronomy at the Johns Hopkins University (Baltimore, MD) for 
his kind hospitality.  

{\bf REFERENCES}               

Fekel et al. 2003, ASPC 303, 113 

Godon, P., Regev, O., \& Shaviv, G., 1995, MNRAS, 275, 1093 

Gonz´alez-Riestra,R., Viotti, R., Iijima, T., \& Greiner, J. 1999, A\&A 347, 478

Gonz´alez-Riestra,R., Viotti, R., Iijima, T., \& Greiner, J. 2008, A\&A 347, 478

Greiner, J. et al. 1997, A\&A, 322, 576

Hubeny, I.1988, Comput.Phys. Commun., 52, 103

Hubeny, I., and Lanz, T.1995, ApJ, 439, 875

Mikolajewska et al. 2003, ASPC 303, 478, 

Mikolajewska, J. et al.1995, AJ, 109, 1289

Muerset et al. 1991, A\&A, 248, 458

Popham, R., \& Narayan, R. 1995, ApJ, 442, 337 

Schmid \& Schild 1997, A\&A 321, 781

Sion, E.M, Acierno, M., \& Turnshek, D.1978, ApJ, 220, 636

Sion, E.M., \& Starrfield, S.G.1994, ApJ, 421, 261 

Skopal, A.2005, A\&A, 440, 995

Smith, V. et al. 1996, A\&A 315, 179

Tomov, N., Tomova, M., Ivanova, A. 2000, A\&A 364, 557 

Tomov, N., Tomova, M.2002, A\&A 388, 202

Viotti, R., Greiner, J., \& Gonzales-Riestra, R.1999, Nuc.Phys.B.Proc.Suppl., 96, 40

Wade, R.A., \& Hubeny, I. 1998, ApJ, 509, 350  

Wood, M. 1995, in {\it White Dwarfs, Proceedings of the 9th European 
Workshop on White Dwarfs} Held at Kiel, Germany, 29 August - 1 September 
1994. {\it Lecture Notes in Physics}, Vol.443, edited by Detlev Koester 
\& Klaus Werner. Springer-Verlag, Berlin, Heidelberg, New York, 1995, p.41

Young, P. R., Dupree, A. K., Espey, B. R., Kenyon, S. J.2006, ApJ, 650, 1091

Young, P.R., Dupree, A., Espey, B.R., Kenyon, S., Ake, T.2005, ApJ, 618, 891

\clearpage

\begin{figure}[!ht]
\plotone{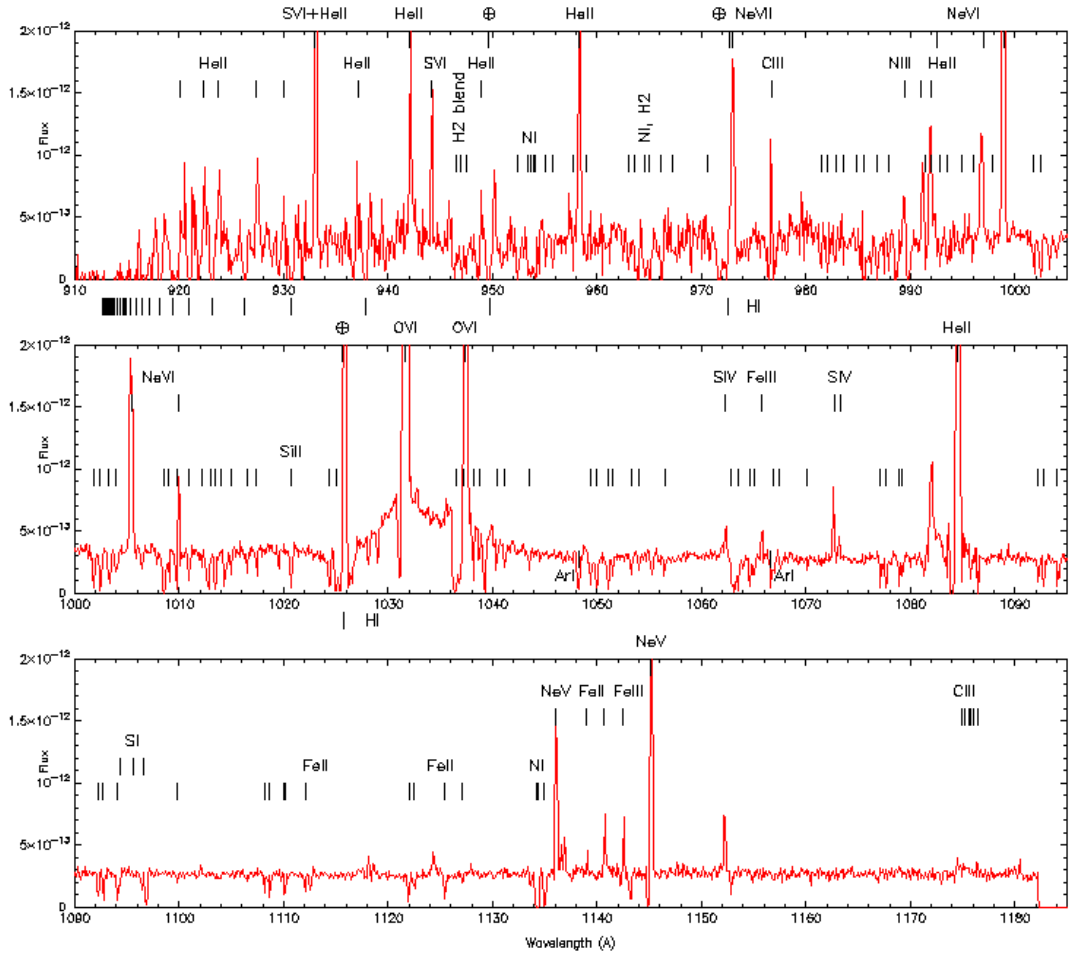}
\caption{Line identifications in the FUSE spectrum of AG Dra,
before it has been dereddened.   
ISM absorption lines have been marked with vertical tick line 
at mid-height in each panel. There are some strong and
sharp emission lines of He\,{\sc ii}, Ne\,{\sc v},{\sc vi} \& {\sc vii} 
and O\,{\sc vi}. Note also the very broad O\,{\sc vi} doublet 
emission feature affecting the hydrogen Ly$\beta$. 
The H\,{\sc i} lines are marked below the upper panel. 
}
\end{figure}

\begin{figure}[!ht]
\plottwo{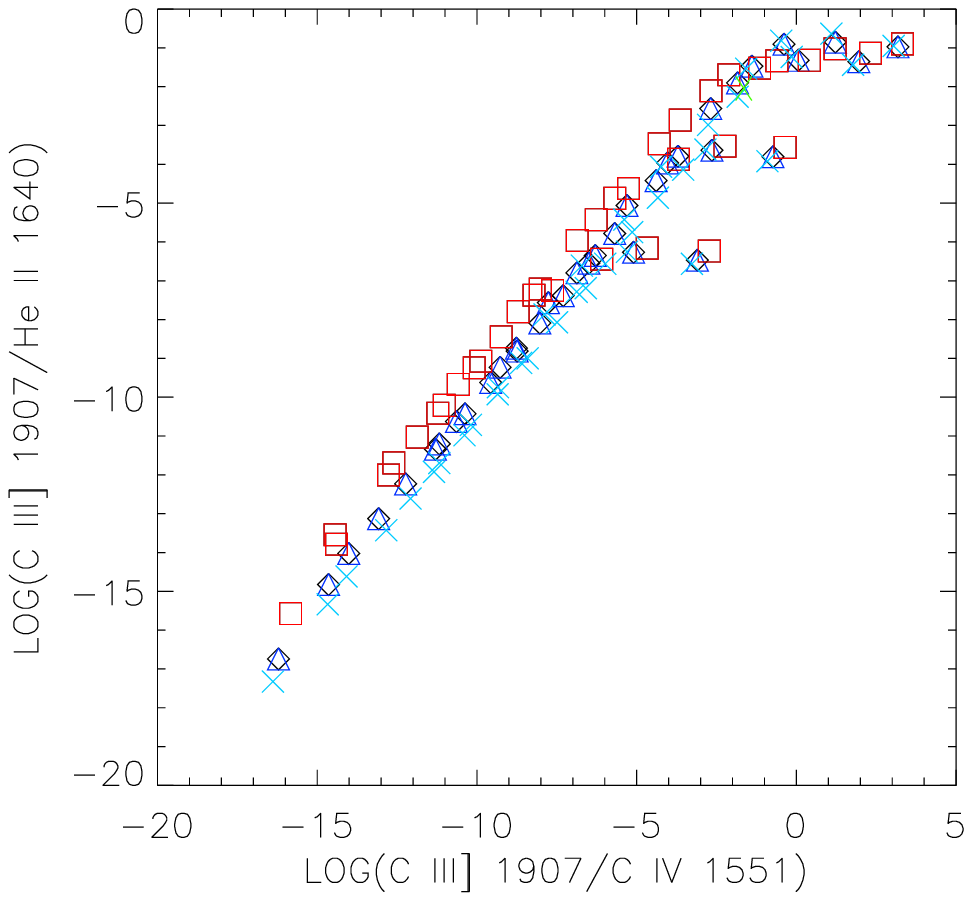}{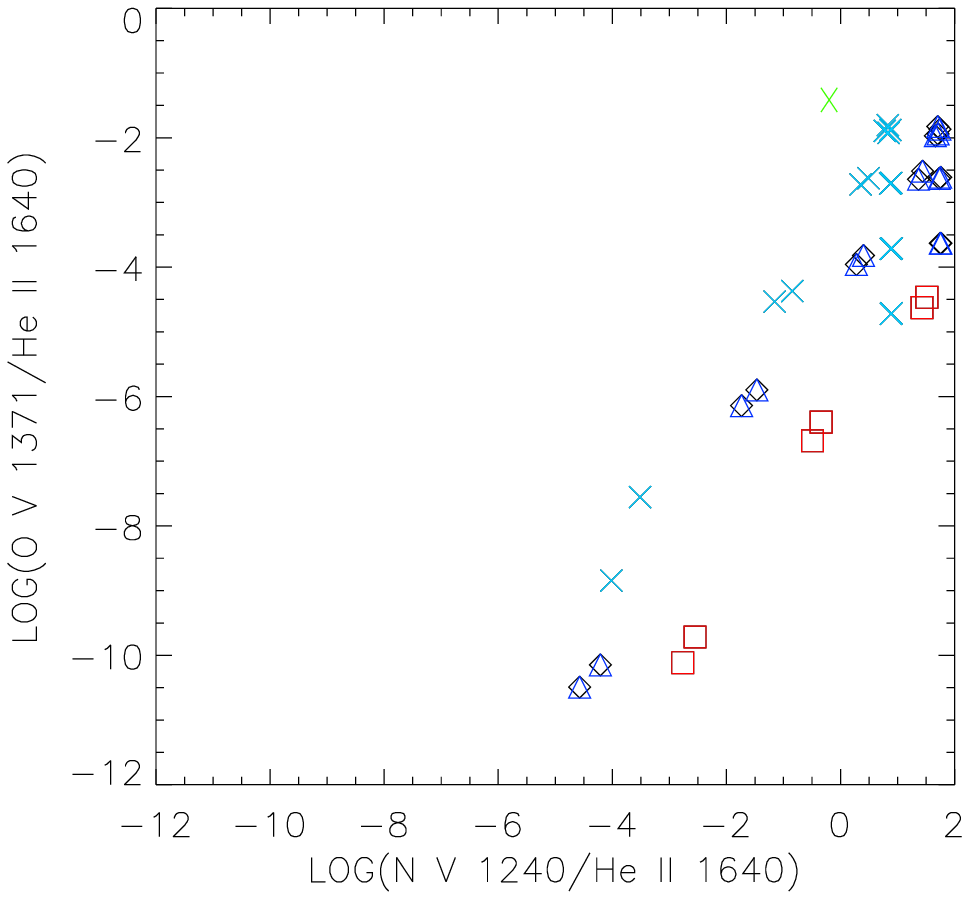}  
\caption{CLOUDY photoionization simulations of AG Dra observed FUV emission line
ratios. The left panel (Fig.2a) displays the behavior of the low ionization line ratios while 
the right panel (Fig.2b) displays the high ionization line ratios. The observed line ratios are denoted by 'X'. 
The simulated line ratios are denoted by red squares, dark blue rhombii, and and light blue 
crosses, and result from incident radiation due to a blackbody at 75,000K, 100,000K, and 
125,000K, respectively.} 
\end{figure}

\begin{figure}[!ht]
\vspace{-12.cm} 
\plotone{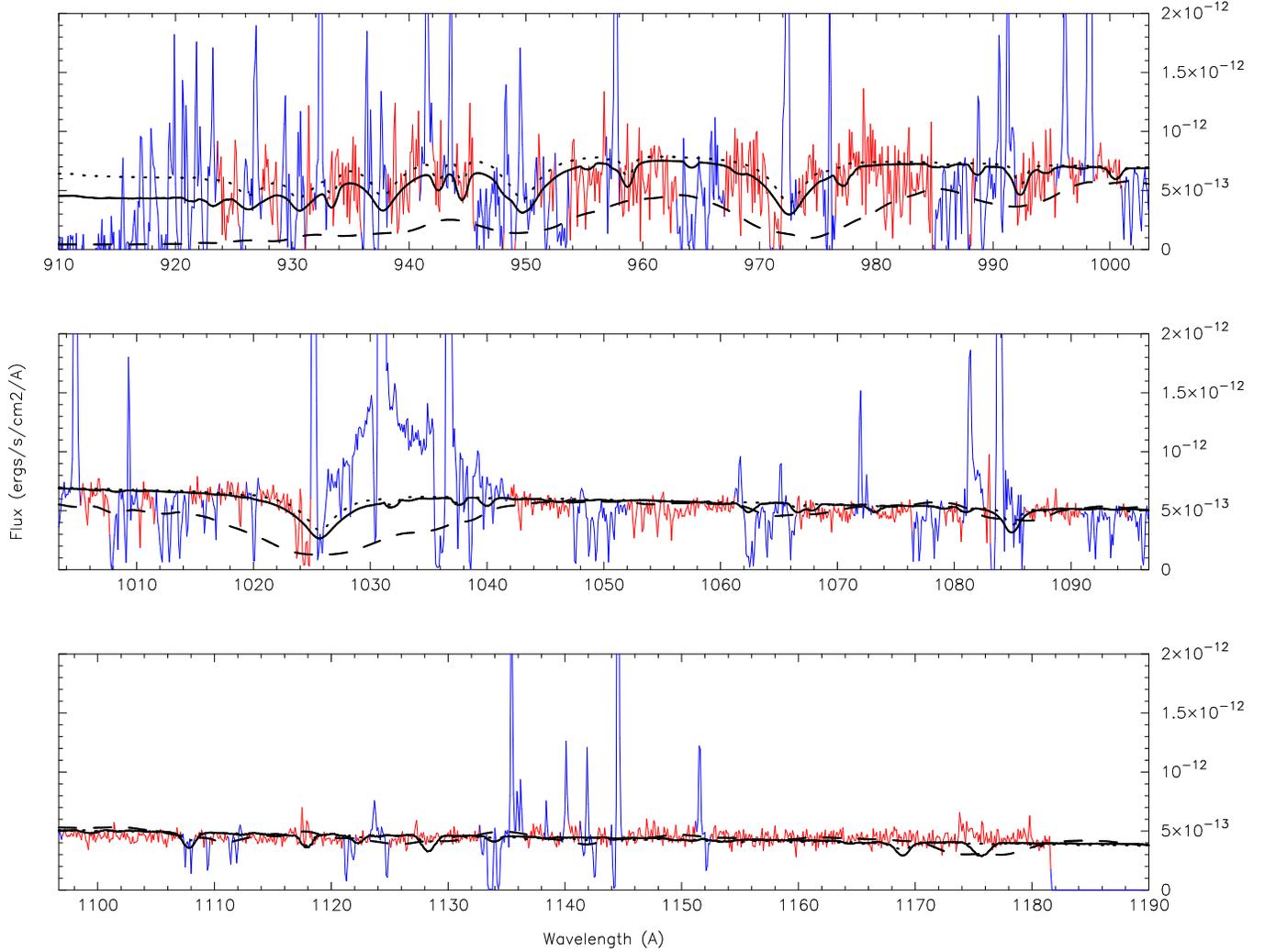}
\caption{We display the three different model fits to the FUSE spectrum. The dashed line is the accretion disk 
fit with $M_{wd}=0.4 M_{\odot}$, $\dot{M} = 5 \times 10^{-8}$ M$_{\sun}$/yr, 
and $i = 41 \degr$, the dotted line is the white dwarf photosphere fit with T$_{eff}$ = 100,000K 
and the solid         line is the white dwarf photosphere fit with T$_{eff}$ = 80,000K. The ISM 
absorption lines and the sharp emission lines have been masked before the fit and are marked in 
blue. We also masked the broad O\,{\sc vi} doublet emission feature.}
\end{figure}

\end{document}